\documentclass[aps,prl,twocolumn,showpacs]{revtex4}%
\usepackage{amssymb}
\usepackage{amsfonts}
\usepackage{graphicx}%
\usepackage{amsmath}%
\usepackage{times}%
\providecommand{\U}[1]{\protect\rule{.1in}{.1in}}
\begin{document}
\title{Role of multiband effects and electron-hole asymmetry in the superconductivity
 and normal state properties of
Ba(Fe$_{1-x}$Co$_{x}$)$_{2}$As$_{2}$}

\author{Lei Fang$^{1}$, Huiqian Luo$^{1}$, Peng Cheng$^{1}$, Zhaosheng Wang$^{1}$,
Ying Jia$^{1}$, Gang Mu$^{1}$, Bing Shen$^{1}$, I. I. Mazin$^{2}$, Lei
Shan$^{1}$, Cong Ren$^{1}$ and Hai-Hu Wen$^{1}$$^{\star}$}
\affiliation{$^{1}$National Laboratory for Superconductivity, Institute of Physics and
Beijing National Laboratory for Condensed Matter Physics, Chinese Academy of
Sciences, P.O. Box 603, Beijing 100190, China}
\affiliation{$^{2}$Code 6391, Naval Research Laboratory, Washington, DC 20375 }

\begin{abstract}
We report a systematic investigation, together with a theoretical
analysis, of the resistivity and Hall effect in single crystals of
Ba(Fe$_{1-x}$Co$_{x}$)$_{2}$As$_{2}$, over a wide doping range. We
find a surprisingly great disparity between the relaxation rates of
the holes and the electrons, in excess of an order of magnitude in
the low-doping, low-temperature regime. The ratio of the electron to
hole mobilities diminishes with temperature and doping (away from
the magnetically ordered state) and becomes more conventional. We
also find a straightforward explanation of the large asymmetry
(compared to cuprates) of the superconducting dome: in the
underdoped regime the decisive factor is the competition between AF
and superconductivity (SC), while in the overdoped regime the main
role is played by degradation of the nesting that weakens the
pairing interaction. Our results indicate that spin-fluctuations due
to interband electron-hole scattering play a crucial role not only
in the superconducting pairing, but also in the normal transport.

\end{abstract}

\maketitle


The discovery in the last year of new iron-based superconductors~\cite{Kamihara} provided a tempting analogy with
high-T$_{c}$ cuprates.  Indeed, a simple
comparison between phase diagrams reveals,  particularly
clearly for the BaFe$_{2}$As$_{2}$
family\cite{Rotter,Sefat,Rotterphys}, a couple interesting
similarities with the cuprates: first and foremost,
 the parent compound is an
antiferromagnet (AFM), and spin fluctuations appear important for carrier
pairing. Second, the superconductivity (SC) appears with either hole or
electron doping, at a finite doping level, and forms a dome-shaped
region in the phase diagram, as in cuprates.

A closer look,
however, reveals equally striking differences:
Indeed, unlike the cuprates, the parent
compoundis in pnictides are metals that support quantum
oscillations\cite{Sebastian,QuantumOs}, and the Coulomb correlations
appear to be
weak\cite{TPD}. Second, unlike cuprates,
superconductivity can be induced without doping,
by external or chemical pressure\cite{Alireza}.
Finally, the
superconducting dome is very asymmetric\cite{JZhao,AJDew}.
And, probably most importantly, electronic structure in cuprates
is formed essentially by one band, while in the pnictiodes
multiband effects are of primary importance.

The doping dependence of the evolution
of the multiband electronic structure
and its relationship to AFM, spin fluctuations, and SC is the key to
the physics of the high-$T_{c}$ ferropnictides.
Systematic Hall coefficient and resistivity measurements are clearly well-suited to
provide useful insight into these issues. In this Letter we
select BaFe$_{2}$As$_{2}$ for a systematic study of the Hall
effect and resistivity. Through quantitative analysis of the experimental
data, combined with theoretical calculations, we establish
a unified view of the doping induced evolution of SC and
AFM, as well as the ramifications for the pairing mechanism.

\begin{figure}[ptb]
\includegraphics[scale=0.73]{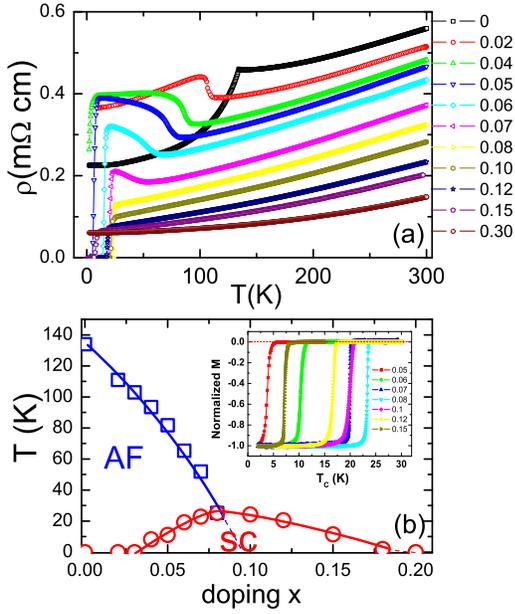}\caption{(Color online) (a)
Temperature dependence of the resistivity of Ba(Fe$_{1-x}$Co$_{x}$)$_{2}$As$_{2}$
single crystals. The maximal $T_{C}$ occurs at about 25.2 K with a doping level of
8\%, where the AF/structural transition cannot be explicitly resolved
(but extrapolates to approximately the same temperature at $x\approx 8$\%). (b)
The phase diagram derived from the resistivity and Hall effect measurements.
We do not resolve the small splitting between
 the structural and AFM transitions.
The inset shows the diamagnetic signal measured in superconducting samples. }%
\label{fig1}%
\end{figure}

The crystals were grown by self-flux method using FeAs as the flux; the
details are described elswhere\cite{JHChuStanford,NNiBaFeCo}. The main
advantage of the 122 system~\cite{Rotter,Sefat,Rotterphys} is that it allows
fabrication of large single crystals, and can be easily doped with
holes\cite{Rotterphys,NNiBaK,HQLuo} or with
electrons\cite{Sefat,GCao,JHChuStanford,NNiBaFeCo}. The crystal structure and
chemical composition were checked by X-ray diffraction and energy dispersive
X-ray microanalysis. The actual concentration of Co in the samples is found to
be close to the nominal values. Measurements of the in-plane longitudinal
($\rho$$_{xx}$) and transverse ($\rho$$_{xy}$) resistances were carried out in
a Physical Properties Measurement System (PPMS, Quantum Design) using the
standard six probe method. All electronic calculations were performed using
the Full Potential LAPW method as implemented in the WIEN2k package. The
experimental crystal structure for BaFe$_{2}$As$_{2}$ was used for all
calculations. Doping was taken into account through the rigid band
approximation. The same setup was used as in Ref. \cite{MazinPRB}.

Fig. \ref{fig1}(a) shows the
resistivity for Ba(Fe$_{1-x}$Co$_{x}$)$_{2}$As$_{2}$ single crystals
with doping levels ranging from the undoped parent phase to heavily
overdoped compounds. With doping, the system evolves from AFM (with
a resistivity anomaly) to superconductivity, and finally to a normal metal.
The resistivity drops sharply at about 137 K in the parent phase,
due to a drastic reduction of the scattering rate in
the AF state that overcomes the reduction of the carrier
concentration due to partial gapping of the
FS~\cite{HuWZoptical,TropeanoOptical}. We have verified by
first-principles calculations that in the fully spin polarized
phase, that is, with the magnetic moment of at least 1.5 $\mu_{B}$,
the calculated Hall concentration is $n_{h}=n_{e}=0.015$, as opposed
to 0.15 in the nonmagnetic case, in quantitative agreement with
the reduction of optical carrier concentration by a factor of 8 and
the relaxation time~\cite{HuWZoptical} by a factor of 20.
Interestingly, a rather small Co doping (2\%) turns this sharp drop
into an equally sharp (though smaller in magnitude) upturn (Fig.
\ref{fig1}(a)). Assuming that the reduction in carrier concentration
is comparable to that at $x=0$, we observe that the reduction of the
relaxation rate in the AFM state is at least 30\% smaller in the
2\%-doped samples than in the undoped ones.  The evolution of the conductivity can be understood as follows:
assuming that the main source of the
transport relaxation is spin fluctuations, the freezing
out of
such fluctuations is more complete when the measured magnetic
moments are larger. Fig. \ref{fig1}(b) shows the phase diagram of Ba(Fe$_{1-x}%
$Co$_{x}$)$_{2}$As$_{2}$ system derived from our data. The characteristic
magnetic anomaly temperature, $T_{AF}$, is determined from resistivity and
Hall coefficient measurements (both sets of data coincide within a few
Kelvin). As mentioned, two points are of interest in regard to this phase
diagram. One is the coexistence/competition between AFM and SC in the
underdoped regime, the other is the asymmetric shape of the superconducting
dome. The Hall measurements presented here provide a comprehensive explanation of both.

\begin{figure}[ptb]
\includegraphics[scale=0.80]{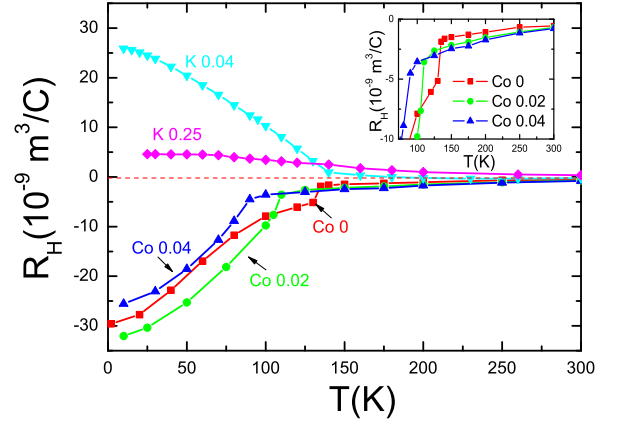}\caption{ (Color online) Temperature dependence of $R_{H}$ for the parent phase, 4\% and 25\% K-doped
Ba$_{1-x}$K$_{x}$Fe$_{2}$As$_{2}$ crystals, 2\% Co-doped and 4\% Co-doped
Ba(Fe$_{2-x}$Co$_{x}$)$_{2}$As$_{2}$ crystals. Note that a very small amount
of K-doping leads to a sudden sign change of $R_{H}$ in the AF state. The
inset shows R$_{H}$ near the AF transition
for selected dopings. }%
\label{fig2}%
\end{figure}

\begin{figure}[ptb]
\includegraphics[scale=0.8]{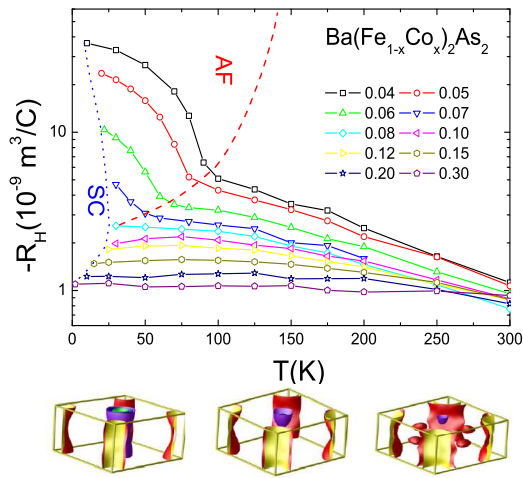}\caption{(Color online) Upper
panel: the temperature dependence of the Hall coefficient $R_{H}$ at
9 T. The red dashed line at $x\lesssim 0.07$ follows $T_{AF}$, below
which $R_{H}$ rises sharply, indicating a dramatic change of the
carrier concentration and scattering rate, as discussed in the text.
The blue dotted line outlines the superconducting region. The lower
panel (from left to right) represents the Fermi surfaces calculated
for nonmagnetic Ba(Fe$_{1-x}$Co$_{x}$)$_{2}$As$_{2}$ for $x=$0, 0.2
and 0.3 (in the
virtual crystal approximation). }
\label{fig3}
\end{figure}

In Figs. \ref{fig2} and \ref{fig3} we present the Hall coefficient $R_{H}$ throughout
wide doping and temperature ranges (until recently, the Hall effect was
measured only at limited doping levels~\cite{JHChuStanford}). The undoped
samples provide the first major surprise. By definition, undoped samples are
compensated, that is, $n_{h}$ = $n_{e}$ = $n_{0}$. The general formula for the
Hall coefficient in the Boltzmann approximation reads~\cite{Ashcroft}
\begin{equation}
R_{H}=\left(  \sum\frac{\sigma_{i}^{2}}{n_{i}^{H}}\right)  /\left(  \sum
\sigma_{i}\right)  ^{2},\label{Eq1}%
\end{equation}
where~$\sigma_{i}=e^{2}(\frac{n}{m})_{i}\tau_{i}$ is the electrical
conductivity in the $i^{th}$ band, $(\frac{n}{m})=\frac{\omega_{p}^{2}}{4\pi
e^{2}}$ is expressed in terms of the plasma frequency for the relevant
crystallographic directions, and $\tau_{i}$ is the Boltzmann relaxation time.
In the case of fully-compensated semimetals, like undoped pnictides, Eq. 1
reduces to

\begin{equation}
R_{H}=n_{0}^{-1}\frac{\sigma_{h}-\sigma_{e}}{\sigma_{h}+\sigma_{e}}=n_{0}%
^{-1}\frac{\mu_{h}-\mu_{e}}{\mu_{h}+\mu_{e}},\label{Eq2}%
\end{equation}
where $\mu=\sigma/n=\tau/m$ is the mobility. Contrary to a common
misconception, the Hall coefficient in compensated or nearly compensated
metals is hardly characteristic of the actual carrier concentrations, but is
substantially reduced, unless one type of carrier has a much higher mobility
than the other. It is therefore most puzzling that in the undoped samples
$R_{H}=-30\times10^{-9}$ m$^{3}/$C, corresponding to 0.013 carrier per Fe.
Since  according to Eq.2, the carrier number $n_{0}^{-1}$ gives the
{\it upper} boundary for $R_{H}$, we are left to conclude that the actual
$n_{0}$ is 0.01 or less, and that the transport is dominated by the electrons.
This conclusion is supported by the concentration dependence of the low-T Hall
coefficient, which reveals a smooth dependence with a sign change around 1.5\%
hole doping and a maximum of 32$\times$ 10$^{-9}$ m$^{3} $/C around 1.5\%
electron doping. This can again be reconciled within the same model, using the
two-band version of Eq. 1:
\begin{equation}
R_{H}=\frac{n_{h}\mu_{h}^{2}+n_{e}\mu_{e}^{2}}{(n_{e}\mu_{e}+n_{h}\mu_{h}%
)^{2}}.\label{Eq3}%
\end{equation}
If $\mu_{e}\gg\mu_{h}$, then $R_{H}\approx1/n_{e}$. However, at the hole
doping with $x$ $\sim$ $n_{0}$ the electron pocket in the AF state
disappears, and the Hall coefficient abruptly changes sign (cf. Fig.
\ref{fig2}).

The normal state Hall data analysis also indicates that electrons
dominate the transport. Just above the AFM transition, $R_{H}$ is
reduced to a value corresponding to a carrier concentration of
0.21 e. Using the calculated nonmagnetic carrier concentration of
0.15 e, we can deduce from Eq. \ref{Eq2} that electron mobility at
that temperature is about 6 times larger than the hole mobility. At
higher temperatures, $R_{H}$ continues to decrease with the
temperature, reaching $-0.56\times10^{-9}$ m$^{3}$/C at $T=300$ K
($-0.56$ e/Fe), corresponding now to $\mu_{h}\sim0.6\mu_{e}$. To
summarize this part, the high-temperature state of
BaFe$_{2}$As$_{2}$ is consistent with the nonmagnetic band structure
calculations, assuming that the hole mobility is smaller than the
electron mobility at room temperature and becomes \textit{much}
smaller upon cooling (essentially negligible at $T\ll T_{AF})$.

Let us now turn to the electron doping.  As explained
above, the sharp increase of $R_{H}$ is gradually less well
expressed with doping, in accordance with the gradual suppression of the
magnetism, but is still detectable in all samples where resistivity
measurements indicate an AF transition (see Fig. \ref{fig3}). As in
the undoped crystals, $R_{H}$ continues to decrease upon heating
above $T_{AF}$, albeit much slower than at low $T$. This indicates
that for $x<0.08$ even at room temperature magnetic fluctuations
still affect the carrier concentration. At higher dopings this
effect disappears, and the temperature dependence becomes rather
moderate and essentially vanishes above $x\approx0.2$, as the hole
pockets practically close. The moderate $T$ dependence at
$0.08<x<0.2$ is readily understood in terms of a somewhat different
$T$ dependence for the hole and electron mobilities.
Importantly, superconductivity vanishes at a doping level of
18--20\%, where the temperature dependence of the Hall coefficient
$R_{H}$ disappears. This indicates that a multiband Fermi surface is a
prerequisite for superconductivity. In the the lower panel of
Fig. \ref{fig3}, we show the Fermi surfaces calculated at $x=$0, 0.2
and 0.3, in the rigid band model. Note that at $x=0.2$ the hole
pockets lose their 2D character (and 2D nesting) and practically
disappear at $x=0.3$.

\begin{figure}[ptb]
\includegraphics[scale=0.8]{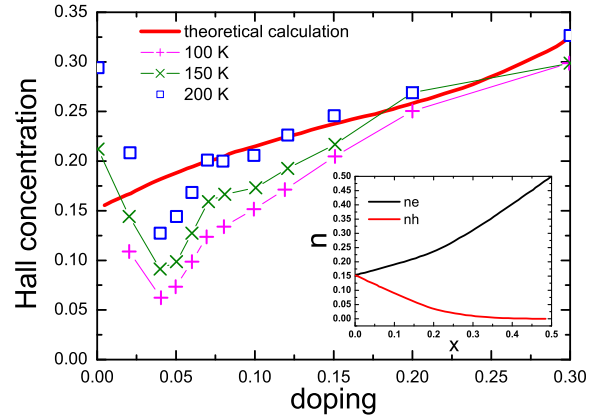}\caption{(Color online) Electron
concentration extracted from the Hall coefficient presented in the
main text, as compared with the calculated Hall concentrations
(solid line), assuming an $x$-dependent ratio of the electron and
hole mobilities, $\mu_h/\mu _e=1.3x$. The inset shows the calculated
volumes of the hole and electron
pockets as a function of electron doping, in the rigid band approximation.}
\label{fig4}
\end{figure}

It is also instructive to analyze the effective Hall concentration as a
function of doping at high temperatures. As Fig. \ref{fig4} shows, the
dependence is non-monotonic, with three distinct regimes: one for
$x\lesssim0.04$, another for $0.04\lesssim x\lesssim0.08$, and the third for
$x\gtrsim0.08.$ In the first regime the effective Hall concentration drops
from a rather large number (twice larger than the calculated nonmagnetic
$n_{0}$, as shown by the solid line) to a number even lower than
$n_{0}$ at $x=0.04$. In the two-band model that can have but one meaning: the
ratio of the hole mobility to the electron mobility sharply decreases with
doping. The fact that even at 200 K the minimal observed Hall concentration is
0.12, smaller than $n_{0}$, indicates that fluctuating spin density waves are
still stealing some carriers even near room temperature. With further doping,
however, this effect rapidly diminishes and at $x>0.07$ the measured
concentration (at 200 K) is consistent with the nominal
concentration calculated from the band structure (see Fig.\ref{fig4}).
Comparung the upper line in the inset graph in Fig. \ref{fig4}. one can
see that the electron-only concentration is close to the 200 K
experimental data at $x>0.07$, but slightly smaller than it even at
$x>0.2$. The difference is accunted for by small, but finite hole
contribution.

As a result, the following picture emerges from our transport
measurements. In the formally stoichiometric, undoped compound at
low temperatures, transport is dominated by electron pockets.
Electron bands already have a higher mobility than the hole bands in
the paramagnetic state, and the relaxation rate for the electrons (but
not as  much for the holes) decreases with cooling, and drops
precipitously below $T_{AF}$. With doping, the gapping becomes less well
expressed. At $x\sim0.04$ there are already enough carriers to
support superconductivity, while at $x\sim0.08$ the gapping
disappears and superconductivity enjoys the full density of states.
In the overdoped regime, $T_{c}$ is controlled by the strength of the
available spin fluctuations. The quality of the quasi-nesting between
the hole and the electron FSs is reduced with doping, and
superconductivity disappears where the hole cylinders disappear at
$x\sim0.2$. In our present picture, we can naturally explain the
asymmetric phase diagram [Fig.\ref{fig1}(b)] because the suppression
of $T_{c}$ is governed by two different mechanisms in the underdoped
and overdoped regimes.

The last corollary, very important for theories striving to explain the
superconducting properties in pnictides, is that the relaxation rates of holes
and electrons are very disparate. This may be the reason for the nonexponential
behavior of such characteristics as penetration depth or the NMR relaxation
rate. Also, this possibility needs to be taken into account when analyzing
optical spectra (as turned out to be the case in MgB$_{2}$\cite{MgB2}). It is
worth noting that in order to explain the temperature dependence of the upper
critical fields in a 1111 compound within a two-band model, one needs at least
an order of magnitude (possibly larger) disparity between the mobilities of the two
bands, even though such an analysis cannot say which band is more
mobile\cite{1111Hc2}. Finally, recent de Haas-van Alphen data also indicate a
higher mobility of electrons~\cite{QuantumOs}.

In summary, we have demonstrated that the superconducting dome in
pnictides has a very different origin from that of the cuprates. The
underdoped side of the dome is defined by competition between AFM
and superconductivity for the carrier
density. In the overdoped regime
superconductivity suffers from a suppression of the spin fluctuations and the loss
of nesting. These two effects together lead to an asymmetric superconducting dome.
Our second result is the surprisingly strong,
and not readily understandable disparity of the scattering rates of the electron
 band and
hole band. \cite{caution}

It is intriguing to ask if these finding are symmetric with respect to the
doping sign. For instance in a hole-doped regime, would the mobility disparity survive,
disappear, or change sign? This question will hopefully be answered by
future experiments.

As a final note, recently similar measurements were reported by
Rullier-Albenque $et$ $al$\cite{FrenchHall}. Their results are very
close to ours, and they also arrive at the conclusion that the only
way to explain the Hall data at low temperatures and low doping is
to assume a drastically suppressed hole mobility (compared to the
electron one). They extend this assumption to all dopings and all
$T< 150$ K (i.e., they assume the one-band model for the entire
range), which we feel is not justified by the data. More
importantly, they did not consider any effects of long-range AFM
fluctuations on the carrier concentration. As a result, they were
forced to introduce a thermally activated behavior for the electron
density, which we believe is unphysical for this system, and
requires not just renormalization of the LDA band structure, but
abandoning it in a qualitative way.

This work was supported by the Natural Science Foundation of
China(973 project No: 2006CB60100, 2006CB921107, 2006CB921802), and
Chinese Academy of Sciences (Project ITSNEM). IIM was supported by
the ONR.

$^{\star}$ hhwen@aphy.iphy.ac.cn


\begin{thebibliography}{99}                                                                                               %
\bibitem {Kamihara}Y. Kamihara \textit{et al.}, {J. Am. Chem. Soc.}
\textbf{130}, 3296 (2008).

\bibitem {Rotter}M. Rotter \textit{et al.}, {Angew. Chem. Int. Ed.}
\textbf{47}, 7949 (2008).

\bibitem {Sefat}A. S. Sefat \textit{et al.}, {Phys. Rev. Lett.}
\textbf{101}, 117004 (2008).

\bibitem {Rotterphys}M. Rotter \textit{et al.}, {Phys. Rev. Lett.}
\textbf{101}, 107006 (2008).

\bibitem {Sebastian}S. E. Sebastian \textit{et al.}, {J. Phys.: Condens.
Matter} \textbf{20}, 422203 (2008).

\bibitem {QuantumOs}A. I. Coldea \textit{et al.}, Phys. Rev. Lett. {\bf 101},
 216402 (2008); J. G. Analytis  \textit{et al.}, Phys. Rev. Lett. {\bf 103}, 076401 (2009).

\bibitem {TPD}W. L. Yang \textit{et al.}, {Phys. Rev.}
\textbf{B 80}, 014508(2009); see also Z. Tesanovic, Phys.
\textbf{2}, 69(2009).


\bibitem {Alireza}P. A. Alireza \textit{et al.}, {J. Phys.: Condens.
Matter} \textbf{21}, 012208 (2009).

\bibitem {JZhao}J. Zhao \textit{et al.}, {\ Nature Materials} \textbf{7},
953 (2008).

\bibitem {AJDew}A. J. Dew \textit{et al.}, {\ Nature Materials} \textbf{8}, 310(2009).

\bibitem {JHChuStanford}J.-H. Chu \textit{et al.}, {Phys. Rev. B }
\textbf{79}, 0145064 (2009).

\bibitem {NNiBaFeCo}N. Ni \textit{et al.}, {Phys. Rev. B }\textbf{\ 78},
214515 (2008).


\bibitem {NNiBaK}N. Ni \textit{et al.}, {Phys. Rev. B} \textbf{\ 78}, 014507 (2008).

\bibitem {HQLuo}H.-Q. Luo \textit{et al.}, {Supercond. Sci. Technol. }
\textbf{21}, 125014 (2008).

\bibitem {GCao}G. Cao \textit{et al.}, {Phys. Rev. B} \textbf{79}, 054521 (2009).

\bibitem {MazinPRB}I. I. Mazin, \textit{et al.,} Phys. Rev. \textbf{B78},
085104 (2008).
\bibitem {HuWZoptical}W. Z. Hu \textit{et al.}, {Phys. Rev. Lett.}
\textbf{101}, 257005 (2008).

\bibitem {TropeanoOptical}M. Tropeano \textit{et al.}, {Supcond. Sci.
Tech.} \textbf{22}, 034004 (2009).

\bibitem {Ashcroft}N. W. Ashcroft, and N. D. Mermin, {Solid State
Physics}, Thompson Learning, Inc. 1976.




\bibitem {MgB2}A. B. Kuz'menko \textit{et al.}, {Solid State Comm.}
\textbf{121}, 479 (2002).

\bibitem {1111Hc2}J. Jaroszynski \textit{et al.}, {Phys. Rev. B.}
\textbf{78}, 174523 (2008).

\bibitem{caution}
As a word of caution, our analysis is based on the two-band Fermi liquid
theory. Strong non-Fermi-liquid effects, such as spin-charge separation, or
strong angular anisotropy of the relaxation rate, may provide alternative
interpretation of our Hall data. We do not see, however, any physical reasons
for either of these effects here.


\bibitem {FrenchHall}F. Rullier-Albenque \textit{et al.}, {Phys. Rev. Lett.}
\textbf{103}, 057001 (2009).
\end{thebibliography}
\end{document}